\documentclass[aps,prx,notitlepage,nofootinbib,floatfix,twocolumn,superscriptaddress, longbibliography]{revtex4-2}

\usepackage{multirow}
\usepackage{graphicx}
\usepackage{bm}
\usepackage{amsmath,amssymb}
\usepackage{hyperref}
\usepackage{lipsum}
\usepackage{color}

\usepackage{bm}
\usepackage{textcase}
\usepackage{textcomp}
\usepackage{booktabs}
\usepackage{braket}
\usepackage{setspace}
\usepackage{mathtools}
\usepackage[markup=default]{changes}
\usepackage{comment}
\definecolor{colorref}{rgb}{0.0, 0.408, 0.647}
\definecolor{grey}{rgb}{0.95, 0.95, 0.95}
\hypersetup{
	colorlinks = true,
	allcolors = colorref
}

\newcommand{\SeeSupply}[1]{Supplemental Information}

\newcommand{\PKU}{\affiliation{2}{School of Physics, Peking University, Beijing 100871, China}}
\newcommand{\IQA}{\affiliation{3}{International Quantum Academy, Shenzhen 518048, China}}
\newcommand{\SUST}{\affiliation{4}{Southern University of Science and Technology, Shenzhen 518055, China}}

\newcommand{\HFNL}{\affiliation{6}{
Shenzhen Branch, Hefei National Laboratory, Shenzhen 518048, China}}

\begin{document}

\title{Measurement-and Feedback-Driven Non-Equilibrium Phase Transitions on a Quantum Processor}

\date{\today}

\author{Zhiyi Wu}
\affiliation{\PKU}\affiliation{\IQA}

\author{Xuandong Sun}
\affiliation{\IQA}\affiliation{\SUST}

\author{Songlei Wang}
\affiliation{\PKU}

\author{Jiawei Zhang}
\affiliation{\IQA}\affiliation{\SUST}

\author{Xiaohan Yang}
\affiliation{\IQA}\affiliation{\SUST}

\author{Ji Chu}
\affiliation{\IQA}

\author{Jingjing Niu}
\affiliation{\IQA}\affiliation{\HFNL}

\author{Youpeng Zhong}
\email{zhongyp@sustech.edu.cn}
\affiliation{\IQA}
\affiliation{\HFNL}

\author{Xiao Chen}
\email{chenaad@bc.edu}
\affiliation{Department of Physics, Boston College, Chestnut Hill, MA 02467, USA}

\author{Zhi-Cheng Yang}
\email{zcyang19@pku.edu.cn}
\affiliation{\PKU}
\affiliation{Center for High Energy Physics, Peking University, Beijing 100871, China}

\author{Dapeng Yu}
\affiliation{\IQA}\affiliation{\PKU}
\affiliation{\HFNL}

\date{\today}

\begin{abstract}

Mid-circuit measurements and feedback operations conditioned on the measurement outcomes are essential for implementing quantum error-correction on quantum hardware. When integrated in quantum many-body dynamics, they can give rise to novel non-equilibrium phase transitions both at the level of each individual quantum trajectory and the averaged quantum channel. Experimentally resolving both transitions on realistic devices has been challenging due to limitations on the fidelity and the significant latency for performing mid-circuit measurements and feedback operations in real time. 
Here, we develop a superconducting quantum processor that enables global mid-circuit measurement with an average quantum non-demolition (QND) fidelity of 98.7\% and fast conditional feedback with a 200\,ns real-time decision latency.
Using this platform, we demonstrate the coexistence of an absorbing-state transition in the quantum channel and a measurement-induced entanglement transition at the level of individual quantum trajectories.
For the absorbing-state transition, we experimentally extract a set of critical exponents at the transition point, which is in excellent agreement with the directed percolation universality class. Crucially, the two transitions occur at distinct values of the tuning parameter. Our results demonstrate that adaptive quantum circuits provide a powerful platform for exploring non-equilibrium quantum many-body dynamics.

\end{abstract}

\maketitle

\textit{Introduction.—}
Quantum circuit models have proved to be a powerful framework for exploring many-body non-equilibrium quantum dynamics, from entanglement growth to hydrodynamics~\cite{fisher2023random,PhysRevX.7.031016,Kos2019DualUnitary,PhysRevX.8.021014, PhysRevX.8.021013,Roberts2017Chaos,PhysRevX.8.031058, PhysRevX.8.031057, PhysRevLett.131.220403}, and are naturally implemented on gate-based quantum hardware, including superconducting processors, Rydberg arrays, and trapped ions~\cite{Joshi2020PRLTrappedIonScrambling,johnson2012heralded,Liang2025PRLAnomalousScrambling,Mi2021ScienceScrambling}. Beyond unitary evolution, hardware capabilities for mid-circuit measurements and feedback—essential for error correction~\cite{RyanAnderson2021,krinner2022realizing,Acharya2024QuantumErrorCorrection,Caune2024,Barber2025,Maurer2025}—enable new dynamic phenomena. 
Moreover, adaptive quantum circuits have been shown to enable more efficient quantum state preparation and the execution of certain quantum algorithms~\cite{Corcoles2021,FossFeig2023,Baeumer2024a,Baeumer2024,Baeumer2025}.
Interspersing unitaries with measurements yields measurement-induced phase transitions (MIPT) in the entanglement structure of individual quantum trajectories. As the measurement rate $p$ is increased across a critical value $p_c^{\rm MIPT}$, the system undergoes a transition: for weak measurement rate $p<p_c^{\rm MIPT}$, states remain volume-law entangled, while for strong measurements $p>p_c^{\rm MIPT}$, repeated collapses drive the system to area-law entangled states~\cite{PhysRevB.98.205136,PhysRevB.100.134306,PhysRevX.9.031009,PhysRevB.99.224307,PhysRevB.101.104302,Gullans2020PRXPurification}. 
Although observing this transition requires post-selection of trajectories, experimental signatures have now been reported on multiple platforms~\cite{noel2022measurement,Koh2023,Hoke2023}.

Incorporating feedback conditioned on measurement outcomes further enriches the dynamics~\cite{Iadecola2023a,Ravindranath2023, PhysRevB.109.L020304, Piroli2023, Sierant2023, ravindranath2025free,Morales2024Unsteerable,Hauser2024AdaptiveSymmetryBreaking,Allocca2024StochasticQuantumControl,pokharel2025orderchaosadaptivecircuits,fossfeig2023experimentaldemonstrationadvantageadaptive,Lu2022MeasurementShortcut,Cemin2025}. 
In particular, adaptive circuits can exhibit absorbing-state phase transitions, a paradigmatic form of non-equilibrium criticality. In the absence of additional symmetries, such transitions are generically described by the directed percolation (DP) universality class, with examples ranging from epidemic-spreading models to turbulent liquid-crystal experiments~\cite{Hinrichsen2000,Takeuchi2007DP}. Above a critical rate $p_c^{\rm abs}$, the dynamics drives the system into a target absorbing state on a short timescale, whereas below this threshold the system remains in an active fluctuating phase for exponentially long times~\cite{Ravindranath2023,PhysRevB.109.L020304,Piroli2023,Sierant2023,ravindranath2025free}.
Unlike the MIPT, this transition is visible in the averaged density matrix without post-selection, since the absorbing state is a fixed point of the quantum channel. Theory suggests that the absorbing transition and the entanglement transition are generically distinct, with $p_c^{\rm abs} > p_c^{\rm MIPT}$~\cite{Ravindranath2023}.

Despite the intensive theoretical investigations, experimentally realizing both the entanglement transition and absorbing-state transition in one quantum platform has remained challenging. This is because both transitions require a sufficiently large number of qubits to suppress finite-size effects, as well as high-fidelity gates and low-latency real-time measurement–feedback operations to preserve coherent dynamics. Although a dissipation-driven non-equilibrium phase transition has been previously observed using a trapped-ion quantum simulator with mid-circuit measurements and qubit reuse~\cite{Chertkov2023}, no active classical feedback operation was performed.

\begin{figure*}[t]
    \centering
    \includegraphics[width=\textwidth]{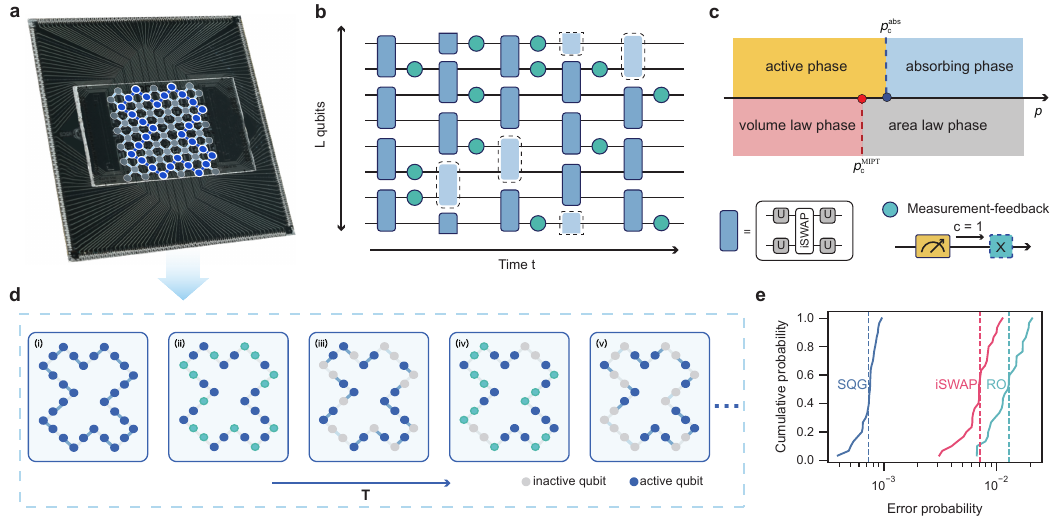}
    \caption{
    \textbf{Quantum processor and adaptive quantum circuit for measurement-feedback dynamics.}
    \textbf{a}, The superconducting quantum processor used in this work. The device hosts 66 transmon qubits arranged in a $6 \times 11$ lattice, with 30 qubits (highlighted in blue) selected for the experiment to minimize crosstalk and enable high-fidelity mid-circuit operations.
    \textbf{b}, Schematics of the adaptive quantum circuit, consisting of alternating layers of two-qubit unitaries and mid-circuit measurements followed by real-time conditional feedback. The two-qubit unitaries are implemented using iSWAP-like entanglers dressed by random equatorial $\pi/2$ phase rotations (denoted by $U$).
    \textbf{c}, Phase diagram illustrating the measurement-and feedback-driven non-equilibrium phase transitions as the measurement rate $p$ is varied.
    \textbf{d}, Illustration for the experimentally executed sequence of the adaptive circuit for the initial state where all sites are occupied. During the evolution, the classical flag variables dynamically track active and inactive qubits under the adaptive feedback protocol. Each labeled step (i, ii, iii, …) consists of a unitary-evolution layer followed by a measurement–feedback layer.
    \textbf{e}, Cumulative error distributions for synchronized SQG, iSWAP, and mid-circuit RO 
    operations, with median error rates of 0.07\%, 0.7\%, and 1.3\%, respectively.
    }
    \label{fig1}
\end{figure*}

Here, we realize both transitions on a superconducting quantum processor with fast mid-circuit measurement and feedback (measurement $\sim 700\,$ns, feedback $\sim 200\,$ns). Using 30 qubits augmented with classical flag variables, we observe a transition between an absorbing phase, where the density of active qubits rapidly vanishes, and an active phase with a finite steady-state activity. 
By varying initial conditions, we extract critical exponents consistent with the DP universality class.
On the same device, using an 8-qubit implementation of the circuit architecture, we also observe a measurement-induced entanglement transition from volume-law to area-law scaling. Crucially, the two critical points are clearly separated, with $p_c^{\rm MIPT} < p_c^{\rm abs}$, demonstrating the distinct physical nature of these non-equilibrium quantum phase transitions.

\begin{figure*}[t]
    \centering
    \includegraphics[width=\textwidth]{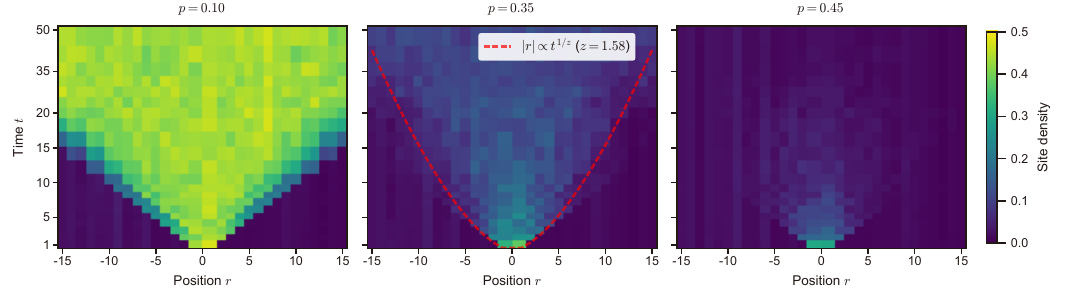}
    \caption{
    \textbf{Absorbing-state transition in the adaptive quantum circuit.}
    Measured spatiotemporal profile of the local occupation $\langle n_i(t)\rangle$
    on the 30-qubit processor for different measurement rates $p$,
    starting from an initial state with a single occupied site in the middle,
    $|\psi_0\rangle = |0\cdots010\cdots0\rangle$.
    Infrequent measurement ($p=0.10$) yields a ballistically spreading active cluster,
    while sufficiently frequent measurement ($p=0.45$) rapidly drives the system into
    the absorbing state with no particle.
    Near the critical point ($p_c^{\rm abs}\simeq0.35$), the spreading becomes
    sub-ballistic with a lightcone satisfying $|r|\sim t^{1/z}$ with a dynamical exponent
    $z = 1.58$, consistent with the directed percolation universality class. (The red dashed line is a guide to the eye.)
    The data are obtained by averaging over 100 random circuit instances and
    10,000 trajectories per circuit.}
    \label{fig2}
\end{figure*}

\textit{Model and implementation on a superconducting quantum processor.-} Our quantum circuit consists of alternating layers of nearest-neighbor two-qubit unitary gates arranged in a brickwork pattern, followed by single-qubit Pauli-$Z$ measurements on a one-dimensional chain of $L$ qubits with periodic boundary conditions, as illustrated in Fig.~\ref{fig1}b.  The unitary layers alternate between odd and even bonds, such that each gate acts only on odd (or even) links within a given layer. For convenience, we interpret the local computational basis state $|1\rangle$ as an occupied site (one particle) and $|0\rangle$ as empty. We take the fully empty configuration $|\psi_t\rangle = |00\cdots 0\rangle$ as our target absorbing state. That is, the dynamics drives the system towards $|\psi_t\rangle$ but cannot leave it.

During each measurement step, if the outcome is $|0\rangle$, no action is taken; if the outcome is $|1\rangle$, a corrective Pauli-$X$ gate is applied to that qubit. This feedback operation steers the system toward the target state $|\psi_t\rangle$, whereas the unitary gates can locally drive it away and do not preserve $|\psi_t\rangle$.
To enable an absorbing-state phase transition, we introduce a classical flag variable $f_i$ for each qubit~\cite{SM}, taking values $f_i = 1$ (active) or $f_i = 0$ (inactive)~\cite{Sierant2023, Piroli2023}. The flags are initialized according to the chosen initial state. After each measurement (and possible feedback), the measured qubit’s flag is set to $f_i = 0$. A two-qubit unitary is applied on bond $(i, i+1)$ only if at least one of the two qubits is active, after which both flags are reset to $f_i = f_{i+1} = 1$. When both are inactive ($f_i = f_{i+1} = 0$), no unitary is applied, leaving the local subspace $|00\rangle$ invariant.

This hybrid quantum circuit, augmented by the flag variables, exhibits an absorbing-state transition in the quantum channel and a measurement-induced entanglement transition at the level of individual quantum trajectories (Fig.~\ref{fig1}c). Alternative approaches to realizing such transitions have been proposed in Refs.~\cite{Ravindranath2023, PhysRevB.109.L020304}, where the unitaries are constrained to have a block structure that preserves a specific invariant subspace. In this work, we focus on the flag-based mechanism, as it is experimentally simpler to implement.

The experiment is performed on a superconducting quantum processor comprising 30 frequency-tunable transmon qubits selected from a larger 66-qubit device~\cite{huang2025exact}, arranged in a one-dimensional chain with periodic boundary conditions, as illustrated in Fig.~\ref{fig1}a. On this platform, we implement an adaptive hybrid circuit composed of alternating nearest-neighbor entangling layers and probabilistic measurement--feedback layers (Fig.~\ref{fig1}b,d). Each unitary layer is realized using entangling gates operated near the iSWAP point, dressed with local single-qubit rotations to generate locally randomized dynamics. After each unitary layer, active qubits are independently selected with probability $p$ for projective $Z$-basis measurements. The measurement outcomes are processed in real time by an FPGA-based controller, and a conditional $X$-reset pulse is applied when the outcome corresponds to the $\ket{1}$ state (classical bit $c=1$ in Fig.~\ref{fig1}b), completing the measurement--feedback cycle.
Representative performance metrics of the single-qubit gates (SQG), iSWAP gates, and mid-circuit readout (RO) are shown in Fig.~\ref{fig1}e, with typical fidelities of $\sim$99.9\%, 99.3\%, and $\sim$98.7\%, respectively. These performance levels enable faithful implementation of the intended hybrid circuit dynamics on the superconducting processor. Further experimental details are provided in the Supplementary Material (SM).

The absorbing-state transition is characterized by the average particle number $N(t)=\sum_i n_i(t)$,
where $n_i=1$ ($0$) denotes an occupied (empty) site. Since $N(t)$ is a linear observable of the density matrix, its expectation value $\langle N(t)\rangle$ can be directly obtained by averaging over all measurement outcomes, without the need for post-selection. This allows the absorbing-state transition to be probed at the level of the quantum channel.

\begin{figure}[t]
    \centering
    \includegraphics[width=0.48\textwidth]{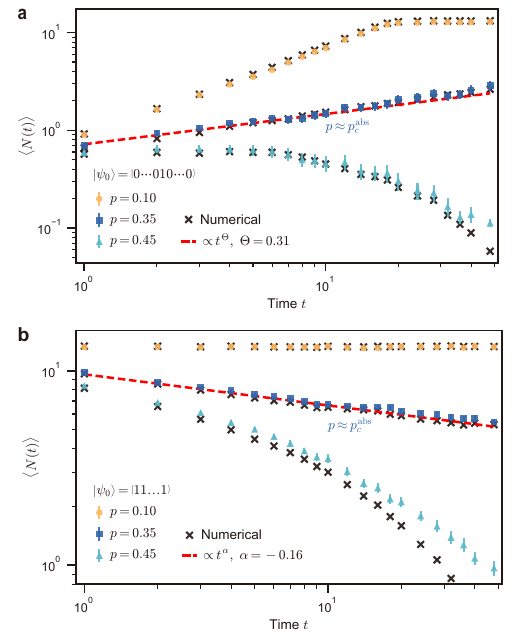}
    \caption{
    \textbf{Critical exponents in directed percolation dynamics.}
    \textbf{a}, Time evolution of the average total particle number \(N(t)\) for the initial state with a single occupied site in the middle $|\psi_0\rangle=|0\cdots010\cdots 0\rangle$.
    \textbf{b}, Time evolution of \(N(t)\) for the fully occupied initial state $|\psi_0\rangle = |11\ldots 1\rangle$.
    In both panels, results are shown for three representative measurement rates:
    \(p=0.10\) (active phase), \(p\approx 0.35\) (critical point), and \(p=0.45\) (absorbing phase).
    Colored symbols represent experimental data (averaged over 100 random circuit instances),
    and black crosses denote ideal numerical simulations with perfect gates and measurements.
    Error bars indicate one standard error of the mean, obtained via bootstrap resampling, and are smaller than the symbol size when not visible. 
    Red dashed lines plot the known exponents of the DP universality class and serve as a guide to the eye.}
    \label{fig3}
\end{figure}

\textit{Absorbing-state transition.-} We initialize the system in a single-seed state $|\psi_0\rangle=|0\cdots010\cdots 0\rangle$ and evolve it under the adaptive hybrid circuit described above.
 At each step, two-qubit unitary gates are applied subject to constraints from the classical flags, followed by mid-circuit measurements and possible feedback operations. We evolve the system for a varying circuit depth up to $t_{\rm max}=50$ before reading out all qubits in Pauli-$Z$ basis, which gives a spatiotemporal profile of the particle configurations upon averaging over trajectories and circuit realizations, as shown in Fig.~\ref{fig2}. The experimental data clearly demonstrate two distinct phases at small and large measurement rates, respectively. When $p$ is small ($p=0.1$ in Fig.~\ref{fig2}), the system is in an active phase where the initial seed spreads ballistically and forms a linear lightcone in spacetime. For the system size we experimentally probe, the total particle number saturates around $t\approx 20$, and this finite density of particles persists at later times up to $t_{\rm max}=50$. This is in stark contrast to the case of larger $p$ ($p=0.45$), where the total particle number increases only a little bit before quickly decaying to zero. This is the absorbing phase where the measurement and feedback operations succeed in steering the system towards the target state $|\psi_t\rangle$. At the critical point $p_c^{\rm abs}\approx 0.35$, the particles spread sub-ballistically in time, showing a sub-linear lightcone structure. 
 We find that the front of this spread follows the scaling $|r|\sim t^{1/z}$ with a dynamical exponent $z=1.58$.
 This exponent agrees with that in the DP universality class~\cite{henkel2008non, SM}.

In Fig.~\ref{fig3}, we plot the average total particle number $N(t)$ in the two phases as well as at the transition point. For \(p = 0.10\), the total number of active sites \(N(t)\) initially increases and then saturates to a steady value, characteristic of the active phase. For \(p = 0.45\), the particle number remains approximately constant for a short period before decaying towards zero, with a small residual density (\(\sim 0.1\)) persisting due to experimental imperfections. Interestingly, at the critical point the total particle number exhibits an algebraic growth in time: $N(t)\sim t^{\Theta}$, with an exponent $\Theta = 0.31$.

The experimental data are consistent with numerical simulations of the same circuits executed experimentally.
Remarkably, the exponent $\Theta$ is also in excellent agreement with that in DP~\cite{henkel2008non, SM}, further suggesting that the transition we observed experimentally belongs to the DP universality class.

To further corroborate the nature of this absorbing-state transition, we perform a complementary set of experiments with a different choice of initial state where all sites are occupied $|\psi_0\rangle = |11\ldots 1\rangle$, from which we also extract another critical exponent. As shown in Fig.~\ref{fig3}b, in the active regime (\(p=0.10\)) the particle density remains essentially constant throughout the evolution, characteristic of a sustained active phase. In the absorbing regime (\(p=0.45\)), the number of active sites rapidly decays toward zero, with a small residual density persisting due to experimental imperfections. Near the critical point (\(p_c^{\rm abs} = 0.35\)), the particle number decays algebraically in time as \(N(t) \sim t^{\alpha}\), with an exponent \(\alpha = -0.16\). This critical exponent also shows excellent agreement with predictions from the one-dimensional DP universality class~\cite{henkel2008non, SM}. Taken together, our experimental observations, supported by numerical simulations, clearly demonstrate a measurement- and feedback-driven absorbing-state transition in our quantum device, belonging to the DP universality class.

\begin{figure}[t]
    \centering
    \includegraphics[width=0.48\textwidth]{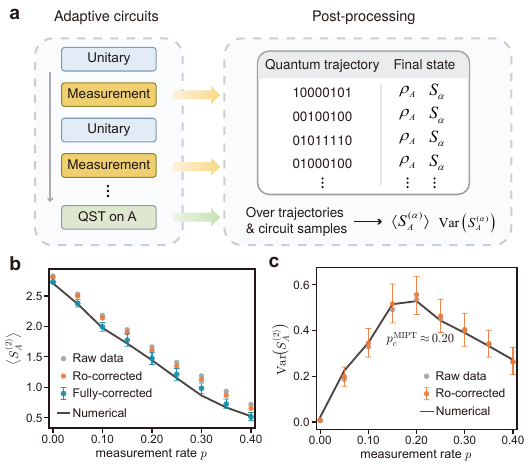}
    \caption{\textbf{Measurement-induced entanglement transition.}
    \textbf{a},  Experimental scheme for trajectory-resolved entanglement characterization. Mid-circuit measurement outcomes are recorded, and for each quantum trajectory the reduced density matrix $\rho_A$ of a three-qubit subsystem $A$ is reconstructed via quantum-state tomography. Post-processing over trajectories and random circuit instances then yields the average second R\'enyi entropy $\langle S_A^{(2)} \rangle$ and its variance $\mathrm{Var}\!\left(S_A^{(2)}\right)$.
    \textbf{b}, Average second Rényi entropy \(\langle S^{(2)}_A \rangle\) versus measurement rate \(p\). Experimental data (dots), corrected for readout errors and residual offsets, 
    agree with ideal numerical simulations (solid line)
    and show a monotonic suppression of entanglement with increasing \(p\).
    \textbf{c}, Variance of $S^{(2)}_A$ \(\mathrm{Var}\!\big(S^{(2)}_A\big)\) across random circuit instances. The variance exhibits a clear peak near \(p \approx 0.20\), which we identify as the measurement-induced entanglement transition. Error bars represent 90\% bootstrap confidence intervals.}
\label{fig4}
\end{figure}

\textit{Entanglement transition.-} To probe the measurement-induced entanglement transition, we implement a reduced eight-qubit realization of the hybrid circuit, which allows full trajectory-level postselection. 
For each quantum trajectory, we reconstruct the reduced density matrix via quantum-state tomography, enabling trajectory-resolved entanglement characterization (Fig.~\ref{fig4}a). 
Since $p_c^{\rm abs}\approx 0.35$, we scan measurement rates $p\in[0,0.4]$ to locate the entanglement transition. 
We quantify entanglement using the second R\'enyi entropy of a contiguous three-qubit subsystem ($|A|=3$) extracted from the final-state reduced density matrix $\rho_A$,
\[
S^{(2)}_A = -\log\!\big(\mathrm{Tr}\,\rho_A^2\big).
\]
Fig.~\ref{fig4}b shows the circuit and trajectory averaged second Rényi entropy \(\langle S^{(2)}_A \rangle\) as a function of the measurement rate \(p\). The result shows a monotonic decrease of entanglement with increasing \(p\), consistent with our expectation.
Since experimental noise introduces a spurious residual classical contribution that biases the raw entropy upward, we implement a two-step error-mitigation procedure. First, tomographic outcomes are corrected using the measured readout calibration matrix (readout inversion). Second, a residual baseline entropy is estimated by measuring the system in the limit \(p=1\) and subsequently subtracted from all data (residual-entropy mitigation). After the above two steps, the error-mitigated experimental data are consistent with numerical simulations within the quoted uncertainties (see SM for a detailed description of the error mitigation procedure~\cite{SM}).

To determine the location of the measurement-induced entanglement transition, we analyze the variance of the second R\'enyi entropy,
\(\mathrm{Var}\!\big(S^{(2)}_{A}\big)
= \langle (S^{(2)}_{A})^{2} \rangle - \langle S^{(2)}_{A} \rangle^{2}\) across randomized circuit instances and quantum trajectories (Fig.~\ref{fig4}c). Based on previous theoretical studies, the variance of $S_A^{(2)}$ develops a pronounced maximum near the critical point and thus provides a diagnostic for the transition~\cite{PhysRevB.100.064204}. After applying the same error-mitigation procedure, the measured variance shows a clear maximum near \(p_c^{\rm MIPT} \approx 0.20\), in agreement with numerical simulations within the experimental uncertainty. This peak signals the location of the entanglement transition. We have also analyzed R\'enyi entropies with different R\'enyi indices $\alpha=1-4$, and the peaks of the variances of all R\'enyi entropies we consider yield a consistent estimate for the entanglement transition point around $p_c^{\rm MIPT} \approx 0.20$~\cite{SM}. 

Importantly, the entanglement transition occurs at a substantially lower measurement rate (\(p_c^{\mathrm{MIPT}} \approx 0.20\)) than the absorbing-state transition extracted from particle-density measurements (\(p_c^{\mathrm{abs}} \approx 0.35\)). This separation demonstrates that the loss of quantum correlations precedes the onset of the macroscopic absorbing-state transition in the feedback-driven circuit. Such a separation of these two transitions was previously suggested theoretically based on numerical simulations of hybrid Haar-random or Clifford circuit models~\cite{Ravindranath2023, PhysRevB.109.L020304, Piroli2023, Sierant2023} and hybrid free fermion systems~\cite{ravindranath2025free}, and to the best of our knowledge, our results provide its first direct experimental observation on a quantum device. Notice that the precise values of \((p_c^{\mathrm{MIPT}}, p_c^{\rm abs})\) depend on the particular gate set used, and are subject to uncertainties arising from the finite subsystem size, the overhead of tomography, and residual experimental noise. In SM, we provide numerical evidence suggesting that the 30-qubit system also enters the area-law entangled phase well before $p_c^{\rm abs}=0.35$~\cite{SM}.

\textit{Summary and outlook.—}
By combining high-fidelity mid-circuit measurements with low-latency conditional feedback, we realize a hybrid circuit architecture that hosts both feedback-driven absorbing-state transitions and measurement-induced entanglement transitions on the same superconducting quantum processor. In a 30-qubit implementation, we observe an absorbing-state transition with critical behavior consistent with the directed percolation universality class, while an 8-qubit realization reveals a separate entanglement transition at a lower measurement rate. The clear separation of the two critical points demonstrates that entanglement suppression and trajectory steering are fundamentally distinct phenomena in adaptive quantum circuits.

More broadly, our work establishes measurement–feedback dynamics as a powerful resource for controlling quantum information flow and exploring stochastic many-body physics on programmable quantum hardware. These capabilities are central to active quantum error correction and adaptive quantum algorithms, and future extensions to larger systems, higher-dimensional geometries, or more complex feedback protocols may reveal new universality classes and dynamical regimes at the interface of quantum information and statistical mechanics.

\textit{Acknowledgements.-}
We thank Libo Zhang and Yuxuan Zhou for the fabrication of the quantum processor.
This work is supported by Grant No. 12375027 from the National Natural Science Foundation of China (S.W. and Z.-C.Y.), the National Natural Science Foundation of China (12374474), the Innovation Program for Quantum Science and Technology (2021ZD0301703), the Science, Technology and Innovation Commission of Shenzhen Municipality (KQTD20210811090049034) and the Guangdong Basic and Applied Basic Research Foundation (2024A1515011714). Numerical simulations were performed on the High-performance Computing Platform of Peking University.

\nocite{Arute2019,Bylander2011,Chen2021,Chow2010,Christandl2012,Emerson2005,Heinsoo2018,Maciejewski2020,Motzoi2009,Nachman2020,Sung2021,Zhang2024,yang2024coupler}

\bibliography{sn-bibliography}

\onecolumngrid
\begin{center}
\bfseries End Matter
\end{center}
\vspace{0.6em}

\noindent\includegraphics[width=\textwidth]    {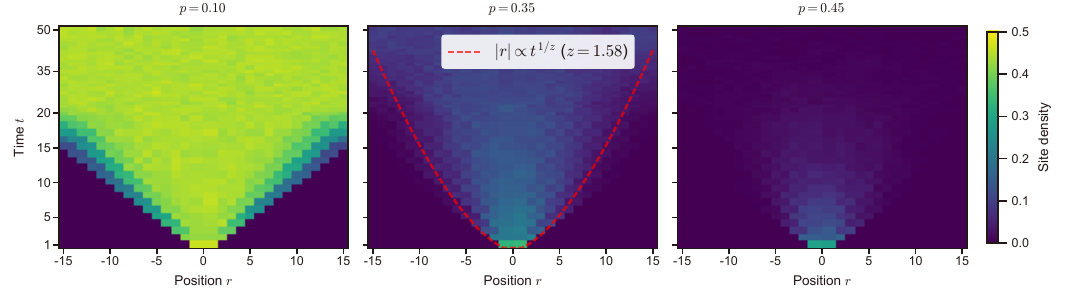}
    \refstepcounter{figure}
    \noindent\textbf{FIG.~\thefigure.}
    \textbf{Numerical simulation of the absorbing-state transition.}
    Spatiotemporal profile of the local occupation $\langle n_i(t)\rangle$
    obtained from tDMRG simulations of the adaptive hybrid circuit,
    corresponding to the experimental data shown in Fig.~2 of the main text.
    The system is initialized in a single-seed state
    $|\psi_0\rangle = |0\cdots010\cdots0\rangle$.
    From left to right, the panels show representative measurement rates
    $p = 0.10$ (active phase), $p \simeq 0.35$ (critical point),
    and $p = 0.45$ (absorbing phase).
    The dashed lines indicate the sub-ballistic spreading
    $|r| \sim t^{1/z}$ at criticality, with $z = 1.58$. (The red dashed line is a guide to the eye.)
    \label{fig:tdmrg_endmatter}
    
    \vspace{0.8em}
    \twocolumngrid
    
    \appendix

To complement the experimental observation of the absorbing-state transition
presented in the main text, we perform time-dependent density-matrix
renormalization group (tDMRG) simulations of the same adaptive circuit protocol implemented in the experiment.
The simulations are initialized in the single-seed state
$|\psi_0\rangle = |0\cdots010\cdots0\rangle$ and follow the identical sequence of
unitary layers, probabilistic measurements, and feedback operations.
The resulting spatiotemporal profiles of the local occupation
$\langle n_i(t)\rangle$, shown in Fig.~\ref{fig:tdmrg_endmatter}, reproduce the characteristic features
of the absorbing-state transition observed experimentally, including ballistic
spreading in the active phase, rapid decay into the absorbing state at large
measurement rates, and sub-ballistic spreading at the critical point.
These numerical results support the interpretation that the experimentally
observed transition belongs to the directed percolation universality class.

\end{document}


\title{Supplementary Information for ``Measurement-and Feedback-Driven Non-Equilibrium Phase Transitions on a Quantum Processor''}

\maketitle

\setcounter{equation}{0}
\setcounter{figure}{0}
\setcounter{table}{0}
\setcounter{page}{1}

\tableofcontents
\newpage

\section{DP universality class}
\label{sec:DP}

In the adaptive quantum circuit studied in the main text, the classical dynamics associated with the quantum channel can be mapped to a stochastic process in which particles (\(A\)) can diffuse, branch (create offspring), or annihilate~\cite{henkel2008non}. The basic reactions are:  
\begin{itemize}
    \item Branching: \(A \to 2A\) \hspace{0.5em} (rate: \(\sigma\))
    \item Annihilation: \(A \to \emptyset\)
\end{itemize}

Depending on the reaction rates, the system exhibits a transition between an \textit{active phase} (a steady state with a finite particle density) and an \textit{absorbing phase} (where all particles eventually disappear). This transition belongs to the directed percolation (DP) universality class. The critical point occurs at a branching rate \(\sigma_c\), which separates the active phase (\(\sigma > \sigma_c\)) from the absorbing phase (\(\sigma < \sigma_c\)). At the critical point (\(\sigma = \sigma_c\)), the system displays scale-invariant dynamics whose properties depend on the initial conditions. Below, we examine two types of initial conditions for a one-dimensional system at criticality.

Two commonly used initial conditions are considered when studying critical dynamics:

\begin{enumerate}
    \item \textbf{Single-Seed Initial Condition:}  
    Starting from a single active site, it can grow into an active cluster in the spatial and temporal direction. The average number of particles grows as a power law in time (at early time):
    \[
    \langle N(t) \rangle \sim t^{\Theta}, \quad \Theta \approx 0.3138.
    \]
    The spatial extent of the active cluster is characterized by the mean-square spreading,
    \[
    \langle R^2(t) \rangle \sim t^{2/z},
    \]
    where \(z \approx 1.581\) is the dynamical exponent relating space and time via \(t \sim L^z\).

    \item \textbf{Fully Active Initial Condition:}  
    When all sites are initially active (or occupied with a nonzero probability \(\rho_0 < 1\)), the particle density decays as
    \[
    \langle \rho(t) \rangle \sim t^{-\alpha}, \quad \alpha \approx 0.1595.
    \]
\end{enumerate}

\section{Classical flag}

\label{app:flag}

The original theoretical proposal of adaptive quantum circuits involves local unitary gates with a block structure, leaving a particular subspace that is locally similar to the target state invariant~\cite{Ravindranath2023, PhysRevB.109.L020304}. Implementation of such unitaries would require a decomposition (compilation) in terms of a set of elementary single-and-two-qubit unitaries that are conveniently executed experimentally. This introduces a huge overhead in terms of the number of gates and circuit depth. To circumvent this problem, we introduce a classical flag variable $f_i$ associated with each qubit that takes value 0 (labeling an `inactive' qubit) or 1 (labeling an `active' qubit), following the method introduced in Ref.\cite{Piroli2023,Sierant2023}. A subsequent entangling gate can only be applied if at least one of the two qubits is active. For the fully active initial state $|\psi_0\rangle = |11\ldots 1\rangle$, we set $f_i=1$ for all $i$ initially. For the single-seed initial state $|\psi_0\rangle=|0\ldots 010\ldots 0\rangle$, we set $f_i=1$ only for the occupied site and $f_i=0$ for all other sites initially.

For each unitary layer, we apply an entangling gate if at least one of $(f_i, f_{i+1})$ is 1. We then reset $f_i=f_{i+1}=1$ after the gate is applied. For each measurement layer, we set $f_i=0$ for the measured qubit. One can easily see that the dynamics of the flag variables themselves satisfy the stochastic classical process described above, and hence exhibit a classical non-equilibrium phase transition in DP universality class. We have also verified this directly in our experiment (data not shown). On the other hand, for the quantum system, since we know that qubits labelled as inactive must be in $|0\rangle$ state whereas those labelled as active typically have a non-zero probability of being in $|1\rangle$ state, we expect that the actual order parameter of the quantum system also inherits the same DP phase transition. Indeed, our results confirm that the quantum order parameter and the classical flag variable exhibit the same DP phase transition under the adaptive dynamics.

\begin{figure*}[t]
    \centering
    \includegraphics[width=0.96\textwidth]{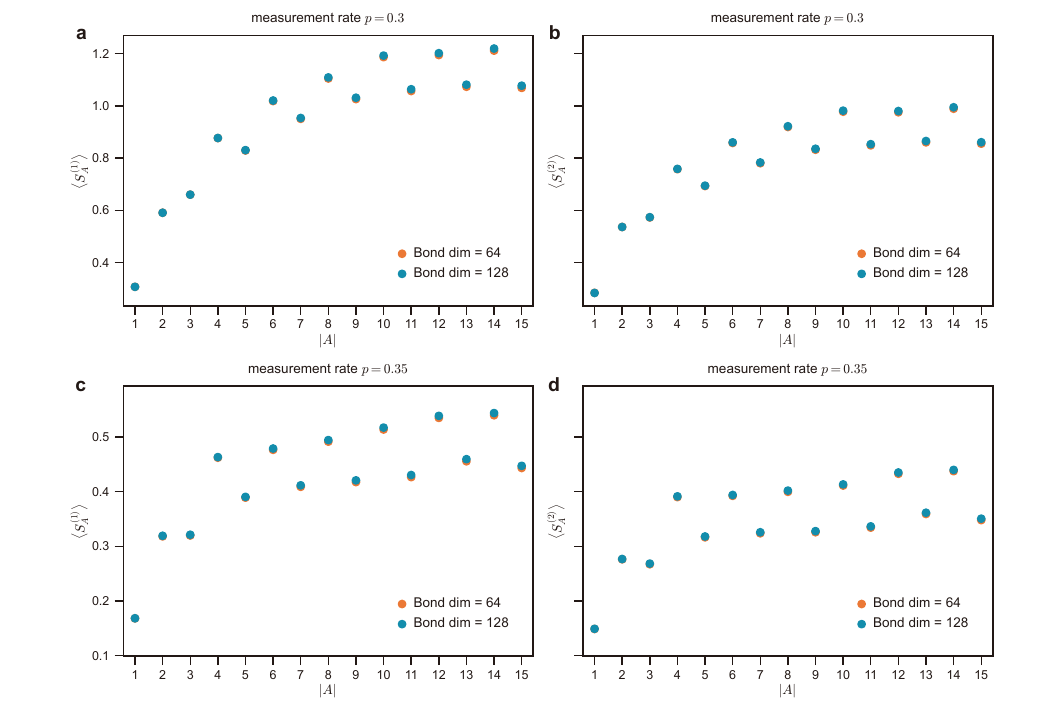}
    \caption{
    \textbf{Scaling of entanglement entropy of the steady states.}
    Numerical tDMRG simulation of the averaged von-Neumann entanglement entropy $\langle S^{(1)}_A \rangle$ (a,c) and the second R\'enyi entropy $\langle S^{(2)}_A\rangle$ (b,d) as a function of subsystem size $|A|$, for the same 30-qubit quantum circuits used experimentally in the main text for probing absorbing-state probingtransition.
    \textbf{a,b,} measurement rate $p=0.3$.
    \textbf{c,d,} measurement rate $p=0.35$.
    The entanglement entropy exhibits area-law scaling already at $p=0.30<p_c^{\rm abs}$. The results for different choices of bond dimension $\chi=64$ and $\chi=128$ are almost identical, confirming numerical convergence.}
    \label{fig:EE_comparison}
\end{figure*}

\section{tDMRG simulation method}

\label{app:simulation}

We numerically simulate the random adaptive quantum circuits described in the main text
under idealized conditions with perfect gates and projective measurements,
using the time-evolving density matrix renormalization group (tDMRG) method. Each quantum trajectory is represented as a matrix product state (MPS), with the bond dimension chosen to ensure numerical convergence. Projective measurements are implemented by first computing the Born probabilities from the occupation numbers of the designated measured qubits and then randomly selecting the subsequent trajectory according to these probabilities. After determining the measurement outcome, we apply the corresponding feedback operation: if the outcome is ``empty,” the MPS is projected onto the empty state; if it is ``occupied,” the MPS is projected onto the occupied state and then subjected to a Pauli-$X$ rotation to reset the qubit to the empty state. After each time step, the MPS is truncated to a maximum bond dimension of $\chi = 64$ and renormalized to compensate for the norm loss introduced by projection. For each circuit instance, we average observables over 50 independent trajectories, and the final results are obtained by averaging these trajectory-averaged quantities over all 500 realizations of the adaptive circuit ensemble. 

In the experiment reported in the main text, the entanglement transition was probed using a smaller system size of eight qubits due to the significant overhead from postselection. Here we provide numerical evidence suggesting that the original 30-qubit system also enters the area-law entangled phase well before the absorbing-state transition at $p_c^{\rm abs}=0.35$. By constructing the reduced density matrix from the MPS, we numerically evaluate the bipartite von-Neumann entropy and the second R\'enyi entropy for the steady states of a system of 30 qubits and subsystem sizes $|A|$ ranging from 1 to 15, as shown in Fig.~\ref{fig:EE_comparison}. The results are averaged over 10 independent trajectories for each of the 500 circuit realizations. We find that indeed the subsystem entanglement saturates for large $|A|$ already at $p=0.30$, showing an area-law scaling. Furthermore, the numerical results for bond dimension $\chi=64$ and $\chi=128$ are almost identical, indicating that a choice of $\chi=64$ is sufficient to ensure convergence for our simulations.

\section{Additional experimental data}

\label{app:data}

\subsection{Fully occupied initial state }
\begin{figure*}[t]
    \centering
    \includegraphics[width=1\textwidth]{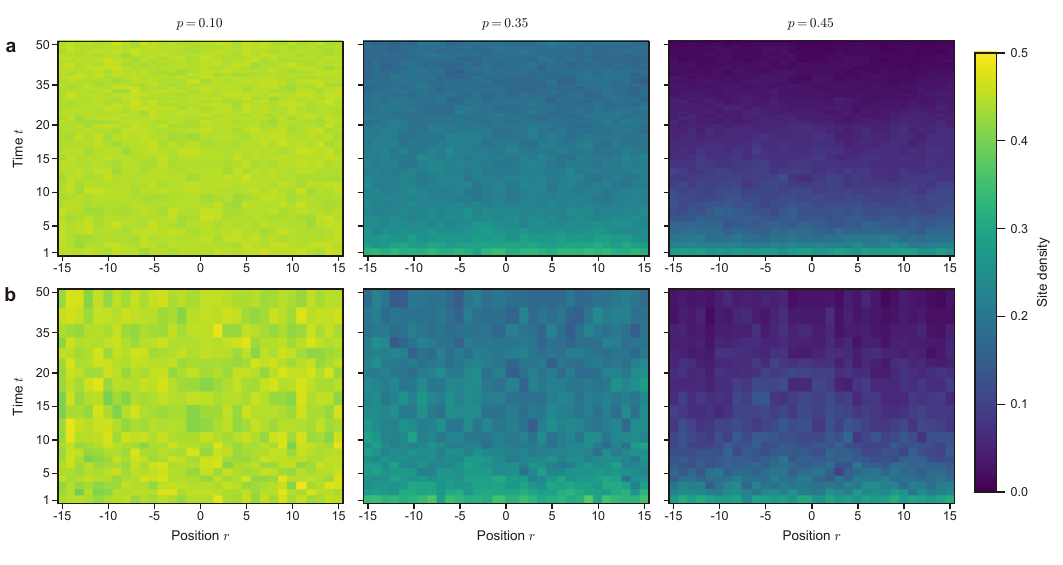}
    \caption{
    \textbf{Absorbing-state transition from fully occupied initial state ($\ket{1}^{\otimes N}$).}
    \textbf{a, Numerical simulation.}
    Spatiotemporal profile of the local occupation $\langle n_i(t)\rangle$ for different measurement rates~$p$ via tDMRG simulations:
    $p = 0.10$ (active phase), $p = 0.35$ (near the critical point), and $p = 0.45$ (absorbing phase), with all qubits 
    initialized in $\ket{1}$. 
    In contrast to the single-site initialization in Fig.~2 of the main text, the dynamics do not exhibit spatial spreading; instead, the 
    occupation decays approximately uniformly in space, reflecting global relaxation induced by repeated measurements and 
    feedback operations.
    \textbf{b, Experiment.}
    Measured spatiotemporal profile of $\langle n_i(t)\rangle$ on the 30-qubit processor under the same protocol. 
    Each pixel represents the average occupation obtained from 100 randomized circuit instances and repeated measurement 
    shots. Color encodes the normalized site density using the same scale as in Fig.~2 of the main text.
    }
    \label{fig_s1}
\end{figure*}

Figure~\ref{fig_s1} shows the absorbing-state dynamics for the fully occupied initial state under the same 
measurement-and-feedback circuit architecture used in the main text. This provides a complementary setup to the one shown in Fig.~2 of the main text starting from a single active site initially.

At low measurement rate ($p = 0.10$), the system remains active, exhibiting nearly uniform occupation throughout the evolution. 
At a near-critical measurement rate ($p = 0.35$), the site density shows a slow, algebraic temporal decay. 
For high measurement rate ($p = 0.45$), the population vanishes rapidly within only a few layers, consistent with the absorbing regime in which repeated measurements and feedback operations drive the system toward the target state exponentially fast. 
Residual late-time signals arise primarily from readout misclassification and occasional leakage events during mid-circuit measurements.
Together with the single active site initial condition in Fig.~2 of the main text, these results establish a consistent picture of an absorbing-state transition in our adaptive circuits.

\subsection{Additional data near the absorbing-state transition}

\begin{figure*}[t]
    \centering
    \includegraphics[width=1\textwidth]{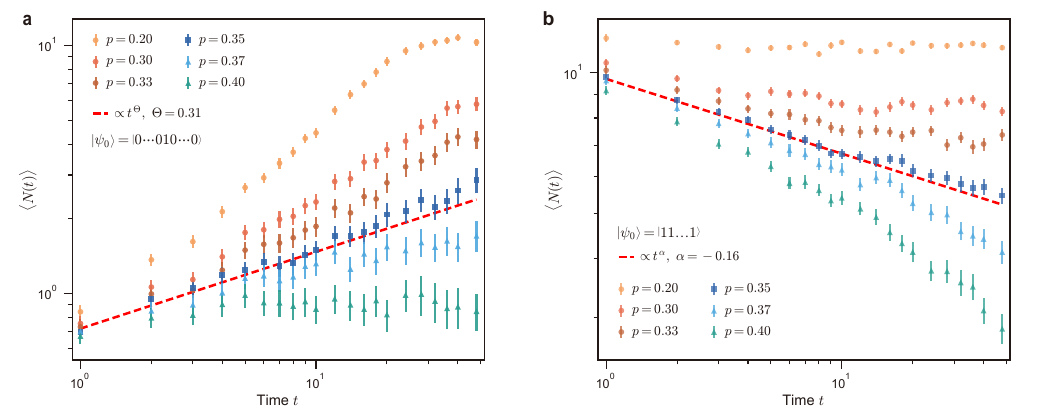}
    \caption{
    \textbf{Additional experimental data near the absorbing-state transition.}
    \textbf{a}, Time evolution of the average total particle number \(N(t)\) for the initial state with a single occupied site in the middle, $|\psi_0\rangle=|0\cdots010\cdots 0\rangle$.
    \textbf{b}, Time evolution of \(N(t)\) for the fully occupied initial state \( |\psi_0\rangle = |11\ldots1\rangle \).
    In both panels, additional measurements are shown for $p=0.30$, $0.33$, $0.37$, and $0.40$, together with the previously reported near-critical point $p\approx0.35$ and the reference point $p=0.20$ near the MIPT critical region.
    The trajectories at $p=0.33$, $0.35$, and $0.37$ remain relatively close at short circuit depths but separate more clearly at later times, with the $p\approx0.35$ data remaining closest to the critical scaling.
    In contrast, the dynamics at $p=0.30$ and $p=0.40$ exhibit more pronounced active- and absorbing-phase behavior, respectively.
    Colored symbols represent experimental data averaged over multiple randomized circuit realizations and repeated measurement shots.
    Error bars indicate one standard error of the mean estimated via bootstrap resampling and are smaller than the symbol size when not visible.
    }
    \label{fig_critical}
\end{figure*}

To better locate the absorbing-state transition point, we perform additional
experiments with a finer scan of $p$ in the vicinity of the presumed critical
region. In addition to the $p=0.35$ data presented in the main text, we further
measure the dynamics at $p=0.30,\,0.33,\,0.37,$ and $0.40$, together with an
additional reference point at $p=0.20$, which lies near the critical region of
the measurement-induced entanglement transition (MIPT).
Figure~\ref{fig_critical} presents the resulting dynamics for two different
initial conditions: a single active site (panel a) and a fully occupied state
(panel b).

For both initializations, the dynamics at $p=0.33$, $0.35$, and $0.37$ remain
relatively close at short circuit depths, indicating a broad near-critical
regime at short times within the correlation time.
Among these values, the trajectories at
$p\approx0.35$ exhibit scaling most consistent with DP criticality,
while the $p=0.33$ and $p=0.37$ data exhibit
slight deviations toward the active and absorbing sides, respectively.
The similarity of these trajectories at short depths suggest a transient critical region
at finite times.
At larger depths, however, the corresponding dynamics become
more clearly separated, which leads to our choice of $p_c=0.35$ as our best estimate for the
absorbing-state transition point within our experimental resolution.

In contrast, the dynamics at $p=0.30$ and $p=0.40$ already exhibit a much more
pronounced separation at short times. The $p=0.30$ data display
characteristics of the active phase, with substantially slower decay and a
more persistent occupation population, whereas the $p=0.40$ trajectories decay
rapidly toward the absorbing regime. The additional reference point at
$p=0.20$ where MIPT happens shows an even clearer active-phase behavior, which demonstrates
a clear separation of these two distinct dynamical phase transitions.

\subsection{Entanglement transition for other R\'enyi entropies}

\begin{figure*}[t]
    \centering
    \includegraphics[width=1\textwidth]{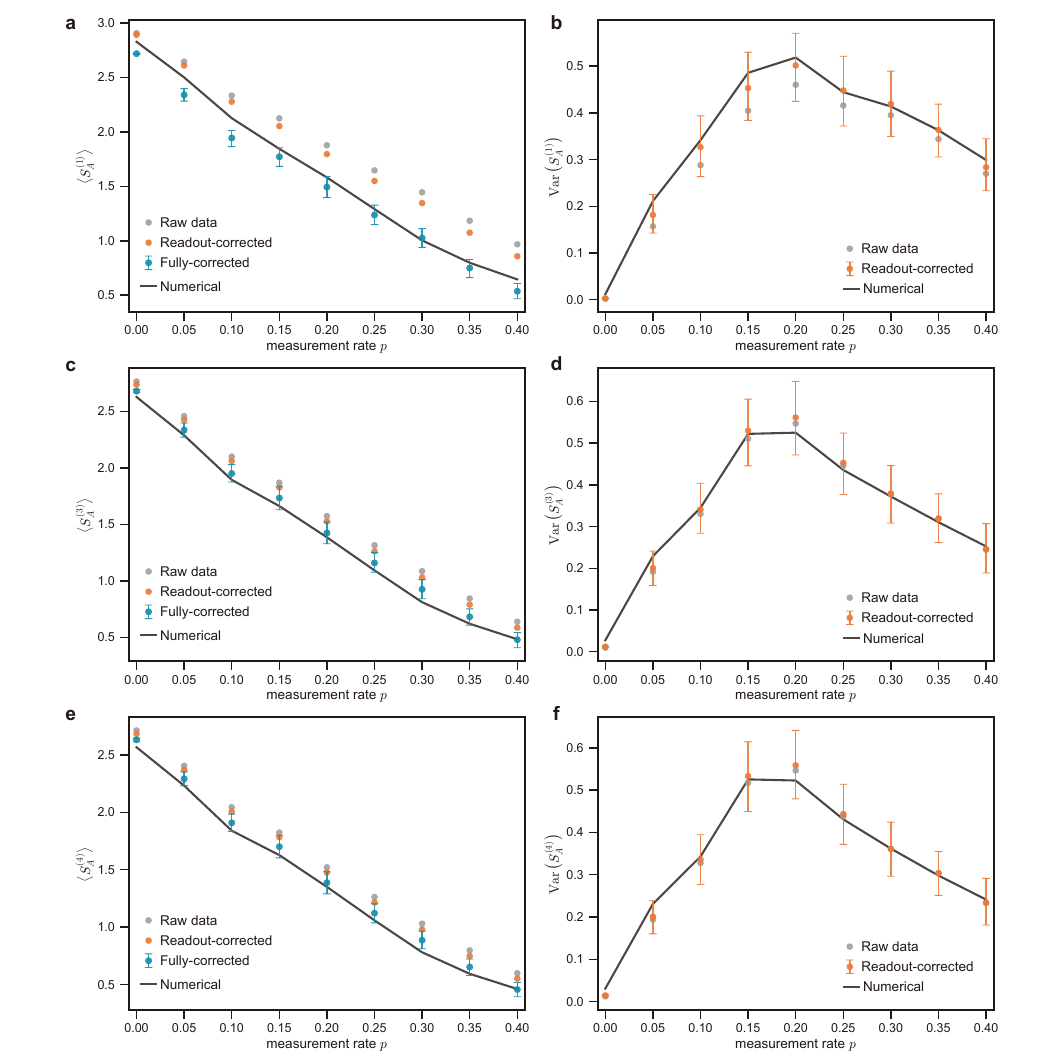}
    \caption{
    \textbf{Entanglement transition at different R\'enyi orders.}
    \textbf{a,c,e,} Average entanglement entropy $\langle S^{(\alpha)}_A \rangle$ for $\alpha = 1,3,4$ as a function of
    measurement rate $p$.  
    \textbf{b,d,f,} Variance of $S^{(\alpha)}_A$ $\mathrm{Var}\!\big(S^{(\alpha)}_A\big)$ for the same orders.  
    All R\'enyi orders exhibit a monotonic decrease with $p$ and a variance peak near a consistent
    entanglement transition point, while quantitative order-dependent differences are discussed in the text.
    }
    \label{fig:s2}
\end{figure*}

In the main text we characterized the measurement-induced entanglement transition using the second 
R\'enyi entropy $S^{(2)}_A$ of a three-qubit subsystem $A$ (Fig.~4b,c).  
Since the full reduced density matrices $\rho_A$ are reconstructed via quantum-state tomography, the same data 
set also allows us to evaluate R\'enyi entropies of other orders $\alpha = 1,3,4$ without additional measurements,
\begin{equation}
    S^{(\alpha)}_A =
    \begin{cases}
       -\operatorname{Tr}\!\big(\rho_A \log \rho_A\big), & \alpha = 1,\\[2pt]
       \dfrac{1}{1-\alpha}\,
       \log\!\Big(\operatorname{Tr}\,\rho_A^{\alpha}\Big), & \alpha \neq 1,
    \end{cases}
\end{equation}
and we apply the same analysis and error-mitigation procedures as for $S^{(2)}_A$.

For all tested orders $\alpha = 1$--$4$, the average entropy $\langle S^{(\alpha)}_A \rangle$ decreases monotonically with 
the measurement rate $p$, while the variance $\mathrm{Var}\!\big(S^{(\alpha)}_A\big)$ exhibits a clear maximum near the 
same critical point $p_c^{\mathrm{MIPT}}\!\approx\!0.20$.  
Although all orders display the same global trend, their detailed behaviors show a clear dependence on the Rényi order.

In particular, the first-order (von Neumann) entropy $S^{(1)}_A$ exhibits a noticeably larger deviation from ideal 
numerics even after full error mitigation.  
This behavior is expected because $S^{(1)}_A$ depends logarithmically on the complete eigenvalue spectrum of $\rho_A$ 
and is therefore especially sensitive to reconstruction errors in the small-eigenvalue sector.  
Similar deviations at $\alpha = 1$—despite correct qualitative scaling—have also been reported in 
previous mid-circuit measurement experiments~\cite{Koh2023}.

In contrast, the higher-order R\'enyi entropies $S^{(3)}_A$ and $S^{(4)}_A$ are less sensitive to contributions from small 
eigenvalues and therefore show much better quantitative agreement with numerics after mitigation, yielding the same 
transition location as $S^{(2)}_A$ within experimental uncertainty. We further note that the peak position of $\mathrm{Var}\!\big(S^{(\alpha)}_A\big)$ displays mild $\alpha$-dependent shifts. 
Nevertheless, the peak locations remain well separated from the absorbing-state transition at 
$p_c^{\mathrm{abs}}\!\approx\!0.35$, confirming that the two transitions are clearly distinct for all R\'enyi orders.

\section{Additional experimental details on entanglement transition}
\label{app:ent_simulation}

\subsection{Simulation setup and parameter selection}

To capture the measurement-induced entanglement transition, we simulate the same hybrid-circuit model used in the experiment reported in the main text and choose circuit parameters that make the transition clearly visible while keeping the sampling cost manageable.  An important choice is the circuit depth: it must be large enough for the subsystem entanglement to reach a quasi-steady value, but not so deep that additional layers only add experimental overhead without altering the 
relevant behavior.

\begin{figure}[t]
    \centering    \includegraphics[width=0.48\textwidth]{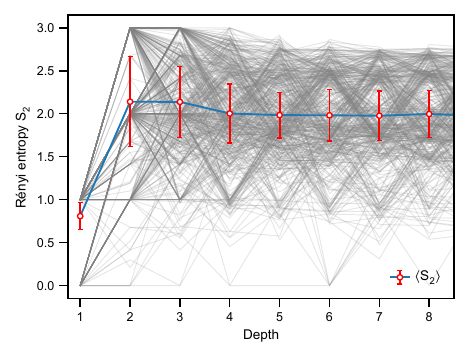}
    \caption{
    \textbf{The second R\'enyi entropy versus circuit depth.}
    Depth dependence of the subsystem R\'enyi entropy $S_A^{(2)}$ at $p=0.1$, obtained from $500$ randomized circuit realizations. Each gray line corresponds to a different random circuit, while the blue line denotes the ensemble average, which quickly settles into a depth-independent value.  The entropy becomes effectively saturated at a depth of about $4$, illustrating the onset of the quasi-steady regime used to select the working depths in the main analysis.
    }
    \label{fig_entropy_depth}
\end{figure}

To determine an appropriate circuit depth, we examine how the R\'enyi entropy evolves with depth across different measurement rates.  As a representative illustration, Fig.~\ref{fig_entropy_depth} shows the depth dependence of the subsystem entanglement entropy $S_A^{(2)}$ at $p=0.1$, evaluated from $500$ randomized circuit realizations.  Each gray line corresponds to a different random circuit, displaying considerable variation at small depth, while the ensemble-averaged entropy (blue line) rapidly settles into a steady trend.  For this and other values of $p$, the entropy becomes essentially depth-independent once the circuit reaches a depth of about $4$, indicating that subsystem entanglement has reached a quasi-steady regime, and that additional layers do not alter the behavior qualitatively. We therefore use depths in the range $1$–$4$ for both simulation and experiment to access this regime without incurring unnecessary sampling overhead.

\subsection{Details on tomography}

Quantum state tomography (QST) ~\cite{Christandl2012} is used to reconstruct the reduced density matrix $\rho_A$ of a small subsystem 
$A$ from an informationally complete set of local measurements.  
For a $k$-qubit block, measuring the three Pauli axes on each qubit yields $3^k$  product bases that uniquely 
determine $\rho_A$ up to statistical fluctuations.  
Once $\rho_A$ is obtained, the R\'enyi entropies $S_A^{(\alpha)}$ for various orders $\alpha$ are directly 
evaluated from its eigenvalues.

In our experiment reported in the main text, we target contiguous three-qubit subsystems ($|A|=3$).  
For each randomized circuit instance at a fixed measurement rate $p$, we measure all $3^3=27$ Pauli-product bases 
and aggregate over shots and instances to reconstruct $\rho_A$.  
This scheme provides full tomographic information while remaining experimentally tractable, enabling the 
extraction of $S_A^{(\alpha)}$ and the corresponding fluctuations across different R\'enyi orders.

\subsection{Sampling strategy and runtime accounting}

To improve data-acquisition efficiency, we first optimize the circuit execution schedule.  
In a conventional setup, each experimental cycle is followed by a long idle interval—typically several multiples of $T_1$, corresponding to an effective reset time of $\sim\!400~\mu\mathrm{s}$—to allow qubits to relax to the ground state $\ket{0}$.  Here we shorten the inter-shot interval by combining an \emph{active reset}—implemented using the same measurement–feedback operation described in the main text, applied once before each circuit to deterministically prepare $\ket{0}$—with a short \emph{passive reset} stage, during which the qubit–coupler configuration is biased to enhance energy dissipation through the readout resonator~\cite{yang2024coupler}. This hybrid reset protocol increases the experimental duty cycle while leaving the in-circuit hybrid dynamics unaffected, reducing the effective cycle time of each circuit to approximately $50~\mu\mathrm{s}$.

With the reset optimization in place, we can sustain the large number of repeated measurements required for QST.  For each value of $p$, we generate about $150$ random circuit instances.  Since extracting $S_A^{(\alpha)}$ requires full QST of a three-qubit subsystem, each circuit is measured in all $3^3 = 27$ Pauli-product bases.  The number of shots per basis is chosen to ensure reliable reconstruction of $\rho_A$: in practice, for most quantum trajectories each tomographic basis is sampled more than $100$ times.

The corresponding circuit depths, median and maximum numbers of mid-circuit measurements, total shots, and overall runtimes for each value of $p$ are summarized in Table~\ref{tab:params}.  The full data-acquisition campaign spans approximately $200$ hours and covers about $1500$ randomized circuit instances across all $p$, providing sufficient statistics to extract the monotonic suppression of $S_A^{(\alpha)}$ with increasing $p$ and to resolve the variance peak near $p_c^{\mathrm{MIPT}}\!\approx\!0.20$, while maintaining stable device performance throughout the experiment.

\begin{table*}[t]
\centering
\setlength{\tabcolsep}{6pt}
\renewcommand{\arraystretch}{1.2}
\begin{tabular}{c c c c c c}
\toprule
$p$ &  Depth & \begin{tabular}{c} Measurements \\ Median (Max)\end{tabular}
    & \begin{tabular}{c}Shots (Approx.)\end{tabular}  &  Random Circuits & \begin{tabular}{c}Wall Time(h)\end{tabular} \\
\midrule
0.00 & 4 & $0(0)$ & $1\times 10^{4}$ & \multirow{10}{*}{\begin{tabular}{c}$\approx 150 \times 10$ \\ $=1500$ total\end{tabular}}
& \multirow{10}{*}{$\approx 200$} \\
0.05 & 4 & $2 (5)$ & $2\times 10^{4}$ & & \\
0.10 & 4 & $3 (7)$ & $5\times 10^{4}$ & & \\
0.15 & 3 & $3 (8)$ & $5\times 10^{4}$ & & \\
0.20 & 3 & $5 (9)$ & $5\times 10^{4}$ & & \\
0.25 & 3 & $6 (10)$ & $1\times 10^{5}$ & & \\
0.30 & 3 & $7 (11)$ & $1\times 10^{5}$ & & \\
0.35 & 3 & $8 (12)$ & $1\times 10^{5}$ & & \\
0.40 & 3 & $8 (12)$ & $1\times 10^{5}$ & & \\
1.00 & 1 & $8 (8)$ & $1\times 10^{5}$ & & \\
\bottomrule
\end{tabular}
\caption{
Experimental parameters and run statistics for the entanglement entropy measurements on hardware.  For each measurement rate $p$, we list the circuit depth, the median (maximum) number of mid-circuit measurements per circuit, the approximate total number of measurement shots, and the number of randomized circuit instances. The rightmost columns indicate that the full data set comprises about $1500$ random circuits and a total wall time of roughly $200$ hours across all values of $p$.
}
\label{tab:params}
\end{table*}

\section{Error mitigation}

\label{app:mitigation}

\subsection{Readout error mitigation.}
In our superconducting platform, raw measurement outcomes exhibit state-dependent detection errors and residual crosstalk between qubits sharing the same readout feedline. To characterize these effects, we calibrate an assignment matrix on the tomographic subsystem by preparing the relevant computational-basis states and measuring them through the same multiplexed readout chain used in the main experiment. This matrix captures the classical stochastic map between the prepared and measured outcomes~\cite{Maciejewski2020,Nachman2020}, including the dominant correlated errors within the subsystem.

In post-processing, we apply a regularized inverse (based on truncated-SVD) to this matrix to obtain corrected estimates of the outcome probabilities prior to state reconstruction. This procedure performs a classical deconvolution of the measurement noise and does not impose any assumptions about the underlying quantum dynamics. All ``readout-corrected’’ data reported in Fig.~4 of the main text and Fig.~\ref{fig:s2} use this correction, while the ``fully corrected’’ results additionally incorporate the residual-entropy procedure described in the next subsection.

\subsection{Residual entropy correction}

After applying readout-error mitigation, the entanglement entropies $S_\alpha(p)$ are extracted from readout-corrected tomographic data that already account for calibrated state-dependent detection errors and readout crosstalk. However, residual discrepancies remain due to incoherent processes such as energy relaxation and dephasing occurring during circuit execution, despite the use of dynamical decoupling to protect idle qubits during measurement windows~\cite{Chen2021,Bylander2011}.
 These effects manifest as an additive positive offset in the entropy that does not originate from genuine many-body entanglement.

To correct for this effect, we follow the linear residual-entropy correction protocol introduced in Ref.~\cite{Hoke2023}. 
The key idea is to use the measurement-only configuration ($p=1$) as a reference, where the system should ideally collapse into a fully disentangled product state with zero intrinsic entropy. 
Assuming that the magnitude of this residual contribution scales linearly with an effective total error budget of the circuit, the corrected entropy is expressed as
\begin{equation}
S_\alpha^{\mathrm{corr}}(p)
=  S_\alpha(p)
- 
\frac{\big\langle \mathcal{E}[\mathcal{C}_{p}] \big\rangle}
{\big\langle \mathcal{E}[\mathcal{C}_{p=1}] \big\rangle}
\,  S_\alpha(p{=}1),
\label{eq:residual_entropy}
\end{equation}
where $\langle \mathcal{E}[\mathcal{C}_{p}] \rangle$ denotes the average effective error budget of circuits at measurement rate $p$ evaluated over random circuit instances, and $ S_\alpha(p{=}1)$ is the corresponding entropy extracted from the readout-corrected data at $p=1$. 
The effective circuit error $\mathcal{E}[\mathcal{C}_p]$ is estimated from the calibrated 
single-qubit, two-qubit, and readout error rates $(\epsilon_{1q}, \epsilon_{2q}, \epsilon_{ro})$, 
weighted by the average number of such operations appearing in the circuit at a given measurement rate~$p$.
In our experiment, the calibrated error rates are 
$\epsilon_{1q} \approx 7\times 10^{-4}$,
$\epsilon_{2q} \approx 7\times 10^{-3}$,
and $\epsilon_{ro} \approx 1.4\times 10^{-2}$,
which are used as the input parameters for this error-budget estimation.
It can be approximately expressed as
\[
\mathcal{E}[\mathcal{C}_p] \approx
\epsilon_{1q}\, N_{1q}(p)
+ \epsilon_{2q}\, N_{2q}(p)
+ \epsilon_{ro}\, N_{ro}(p),
\]
where $N_{1q}(p)$, $N_{2q}(p)$, and $N_{ro}(p)$ denote the mean counts of one-qubit gates, two-qubit gates, and measurement operations, respectively, in a single realization of the circuit at measurement rate~$p$. The averaging $\langle \cdot \rangle$ over different random circuit instances is then performed to obtain $\langle \mathcal{E}[\mathcal{C}_p]\rangle$, which captures the overall error level corresponding to each experimental configuration. The reference value $ S_\alpha(p{=}1)$ is directly measured on the same device and qubit subset to ensure a consistent calibration of this residual floor.

This linear correction removes the noise-induced entropy offset arising from incoherent processes and yields a more accurate estimate of the mean entanglement entropy across different measurement rates~$p$. While it slightly shifts the absolute values of $\langle S_\alpha \rangle$, it does not affect the entropy fluctuations or the extracted variance $\mathrm{Var}(S_\alpha)$, from which the critical measurement point of the entanglement transition is determined. The fully corrected mean entropies are used in the entropy panels of Fig.~4b of the main text and in the multi-order entropy results of Fig.~\ref{fig:s2}a,c,e, where an accurate absolute entropy scale is required for comparison with numerical simulations.

\section{Feedback calibration and verification}

\label{app:feedback}

\subsection{Real-time feedback architecture}

The experiment reported in the main text employs adaptive quantum circuits in which mid-circuit measurements are followed by real-time classical processing and conditional quantum operations. Each feedback cycle consists of four sequential stages: (i) measurement acquisition, (ii) digitization and demodulation, (iii) real-time classification, and (iv) conditional actuation.

During the measurement acquisition stage, a readout pulse of approximately $500~\mathrm{ns}$ is applied to the qubit’s dispersively coupled resonator.  To reliably extract the measurement outcome, we allocate an additional $200~\mathrm{ns}$ in the demodulation window of the ADC--FPGA chain, accommodating the residual photon leakage of the readout resonator $150~\mathrm{ns}$ and the $50~\mathrm{ns}$ propagation time of the signal through the cables to the digitizer. Once the digitized record is available, the FPGA performs real-time processing of the measurement outcome. In our hardware the minimal achievable feedback latency is about $180~\mathrm{ns}$
~\cite{Zhang2024},
but we employ a conservative fixed delay of $200~\mathrm{ns}$ to synchronize the feedback timing across all 30 qubits. Finally, the conditional reset operation is delivered as a short $50~\mathrm{ns}$ feedback pulse, completing the timing-aligned feedback branch across all qubits.

In total, these stages yield an end-to-end latency of approximately $950~\mathrm{ns}$ for a full measurement–feedback cycle, which is far shorter than the qubit energy-relaxation time ($T_1 \sim 70{-}80~\mu\mathrm{s}$) and therefore compatible with coherent mid-circuit control. A detailed timing budget for both the measurement-only and feedback-enabled branches is summarized in Table~\ref{tab:latency}.

\begin{table}[h]
    \centering
    \caption{Timing budget of the real-time measurement and feedback cycle.}
    \begin{tabular}{lc}
        \toprule
        Process & Duration (ns) \\
        \midrule
        Readout pulse (DAC) & $500$ \\
        Residual demodulation tail & $200$ \\
        FPGA synchronization delay & $200$ \\
        Conditional reset pulse & $50$ \\
        \midrule
        \textbf{Total (measurement only)} & $700$ \\
        \textbf{Total (measurement + feedback)} & $950$ \\
        \bottomrule
    \end{tabular}
    \label{tab:latency}
\end{table}

\subsection{Mid-circuit measurement and feedback fidelity}

\begin{figure}[t]
    \centering
    \includegraphics[width=0.48\textwidth]{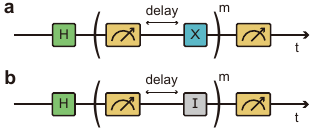}
    \caption{
    \textbf{Circuit used to calibrate the QND mid-circuit measurement fidelity.}
    \textbf{a}, The $X$ gate implements the $\pi$-QND configuration used for the $\ket{1}$-state readout.
    \textbf{b}, Replacing this rotation with an identity operation yields the I-QND configuration for the $\ket{0}$-state readout.}
    \label{fig_qQND}
\end{figure}

We first characterize the fidelity of mid-circuit measurements, which serves as the baseline for evaluating the performance of feedback operations.  
As illustrated in Fig.~\ref{fig_qQND}, the calibration is based on a quantum-nondemolition (QND) protocol that permits repeated measurements of the same qubit without disturbing its state.  
Depending on whether the two measurements are performed consecutively (``I-QND'') or with an intervening $X$ gate (``$\pi$-QND''), the protocol characterizes the readout fidelities of the $\ket{0}$ and $\ket{1}$ states, while suppressing the influence of residual thermal excitation and imperfect state preparation.

For the $\ket{0}$ state, we initialize the qubit in $\ket{0}$, perform a QND measurement, and postselect on $m_1 = 0$.  
We then define the ground-state QND fidelity as
$F_{0}^{\mathrm{QND}} = P(m_2 = 0 \mid m_1 = 0)$, 
where $m_1$ and $m_2$ denote the outcomes of the first and second measurements, respectively.  

\begin{figure}[t]
    \centering
    \includegraphics[width=0.5\textwidth]{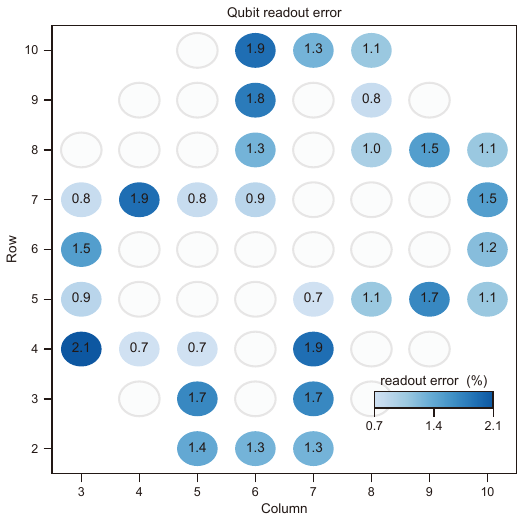}
    \caption{
    \textbf{Readout error.}
    The values are obtained from simultaneous QND mid-circuit measurements performed across all 30 qubits used in the experiment.}
    \label{readout_error}
\end{figure}

For the $\ket{1}$ state, we employ the $\pi$-QND protocol: after postselecting on $m_1 = 0$ from the first measurement, we apply a high-fidelity $X$ gate to prepare $\ket{1}$ and perform a second QND measurement.  
The excited-state QND fidelity is defined as
$F_{1}^{\mathrm{QND}} = P(m_2 = 1 \mid m_1 = 0, X)$
and is mainly limited by energy relaxation during the readout interval.  
The overall mid-circuit measurement fidelity is then
\begin{equation}
    F_{\mathrm{meas}}
    = \frac{F_{0}^{\mathrm{QND}} + F_{1}^{\mathrm{QND}}}{2}.
\end{equation}
Experimentally, $F_{0}^{\mathrm{QND}} \approx 99.5\%$ and $F_{1}^{\mathrm{QND}} \approx 98\%$, yielding an average mid-circuit measurement fidelity of about $98.7\%$.This corresponds to a median readout error of $\sim 1.3\%$, as summarized in Fig.~\ref{readout_error}.

Building upon the QND calibration, we characterize the fidelity of the feedback operation.  
The feedback protocol differs from QND readout only by the presence of a fixed latency before the subsequent gate execution.  
To emulate the real hardware timing, we insert a programmable delay of approximately $200~\mathrm{ns}$ between the mid-circuit measurement and the following operation.  
The ground-state branch is identical to I-QND and we set $F_{0}^{\mathrm{fb}} = F_{0}^{\mathrm{QND}}$.  
For the $\ket{1}$ branch, we use a delayed $\pi$-QND sequence: prepare $\ket{1}$, perform a QND measurement and postselect on $m_1 = 1$, wait for a delay $\tau_{\mathrm{fb}}$, apply an $X$ gate to reset the state to $\ket{0}$, and perform a second measurement.  
The corresponding fidelity is defined as  
$F_{1}^{\mathrm{fb}} = P(m_2 = 0 \mid m_1 = 1, \tau_{\mathrm{fb}}, X)$, 
and the overall feedback fidelity is
\[
    F_{\mathrm{fb}} = \frac{F_{0}^{\mathrm{fb}} + F_{1}^{\mathrm{fb}}}{2}.
\]

Experimentally, the ground-state branch remains unchanged, while the $\ket{1}$-branch feedback fidelity is about $97.3\%$, consistent with the expected relaxation over the calibrated delay.  
This yields an overall feedback-operation fidelity of approximately $98.4\%$.

\section{Superconducting quantum processor and calibration}
\label{app:calibration}

\subsection{Experimental platform}

The experiments are performed on a 66-qubit superconducting quantum processor based on frequency-tunable transmon qubits. Detailed information on device fabrication, chip architecture, and packaging can be found in Ref.~\cite{huang2025exact} (see its Supplementary Material, Sec.~IV).

\begin{figure*}[t]
    \centering
    \includegraphics[width=1\textwidth]{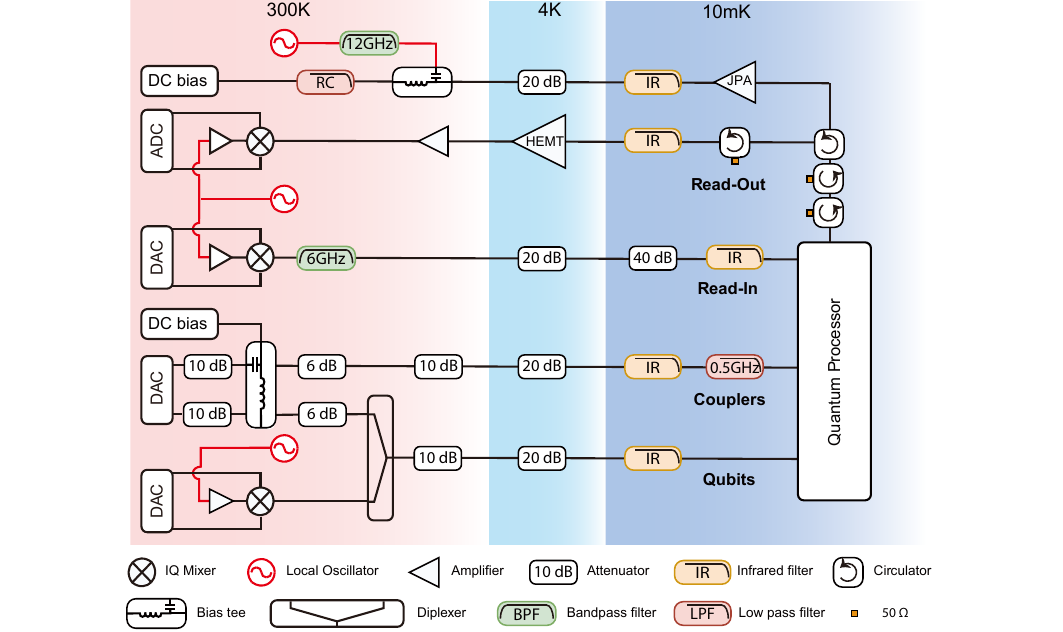}
    \caption{
    \textbf{Experimental setup.}
    Schematic of the cryogenic wiring and room-temperature electronics for control, biasing, and readout of the quantum processor in a dilution refrigerator.}
    \label{wiring}
\end{figure*}

The processor is mounted on the mixing chamber stage of a dilution refrigerator with a base temperature of approximately 10 mK. A schematic of the measurement setup, including cryogenic wiring and room-temperature electronics, is shown in Fig.~\ref{wiring}. Qubit control and readout are implemented using a microwave measurement and control system (M2CS) with integrated DACs and ADCs~\cite{Zhang2024}. Flux biasing for qubits and tunable couplers is applied via dedicated lines combining DC and fast flux pulses through bias-tees. Readout signals are first amplified by Josephson parametric amplifiers (JPAs) at base temperature, followed by high-electron-mobility transistor (HEMT) amplifiers at 4 K and further amplification at room temperature. All lines are equipped with attenuation, filtering, and isolation stages to suppress thermal noise and electromagnetic interference.

\subsection{Qubit selection and layout}

In this processor, the qubits form a $6\times 11$ rectangular lattice of frequency-tunable transmons with nearest-neighbor coupling mediated by tunable couplers~\cite{huang2025exact}. 
Along each vertical column, six qubits share a common readout resonator coupled to a single feedline, enabling frequency-multiplexed dispersive readout. 
Simultaneous readout within the same group can introduce correlated errors due to spectral crowding, amplifier nonlinearity, and residual crosstalk between qubits on the same feedline~\cite{Heinsoo2018}.

For the measurement--feedback experiments, we select 30 qubits from the full device, forming an X-shaped one-dimensional chain embedded in the two-dimensional lattice, as shown in Fig.~1a of the main text. 
This geometry allows the effective one-dimensional topology required by the protocol to wrap around the device while maximizing the spatial separation between qubits sharing the same readout resonator. 
As a result, active qubits are distributed across different readout groups, and qubits coupled to the same resonator are placed as far apart as possible on the lattice. 
This selection strategy reduces the loading on each readout line and minimizes correlated readout errors, thereby enabling high-fidelity simultaneous mid-circuit measurements while maintaining the required connectivity for the experimental protocol.

\subsection{Qubit parameters and frequency allocation}

\begin{figure*}[t]
    \centering
    \includegraphics[width=1\textwidth]{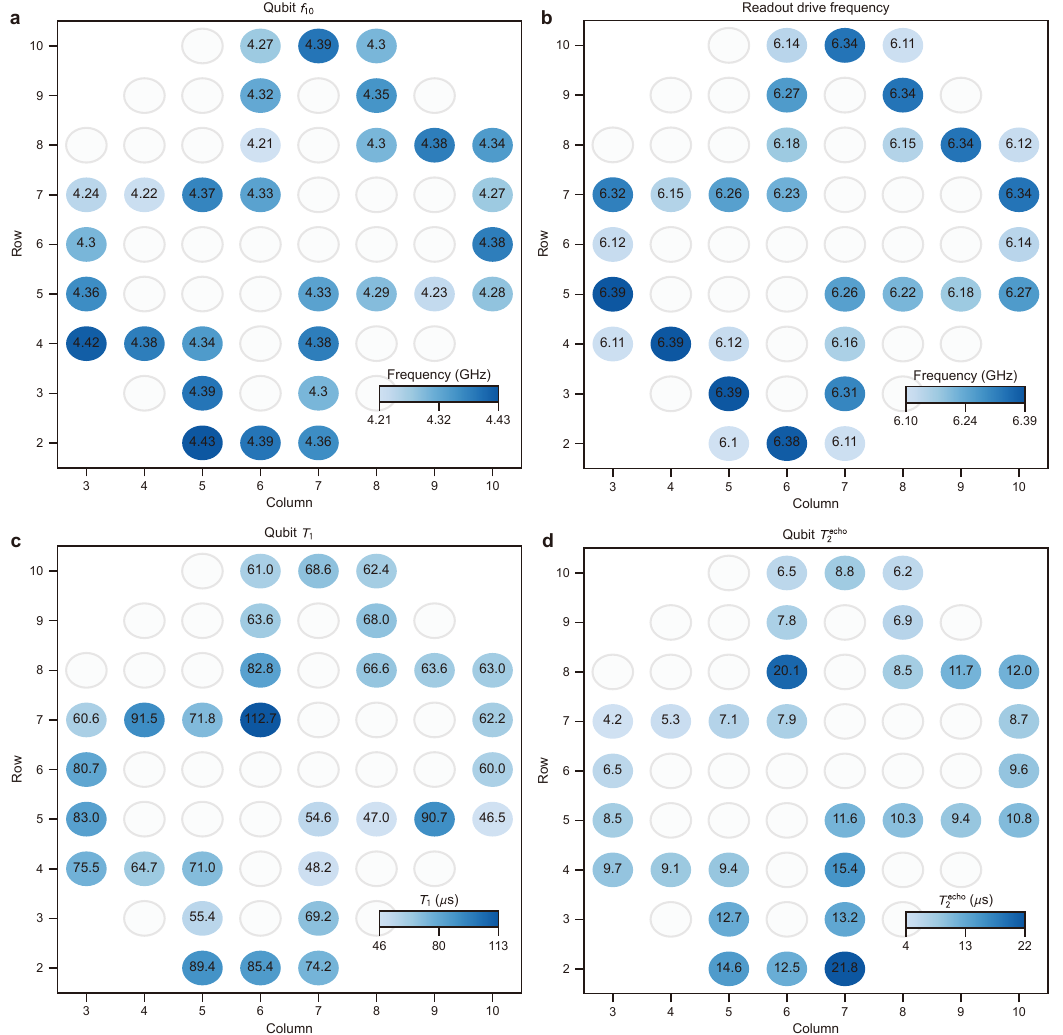}
    \caption{
    \textbf{Qubit parameter map for the 30 qubits used in the experiment.}
    Data are taken from a 66-qubit superconducting processor arranged in a $6\times 11$ lattice. For visualization, the lattice is shown in a $45^\circ$ rotated orientation, where groups of six diagonally aligned qubits share a common readout line.
    \textbf{a}, Operating qubit frequencies $f_{\mathrm{q}}$ obtained from spectroscopy.
    \textbf{b}, Calibrated readout-drive frequencies of the corresponding resonators.
    \textbf{c}, Energy-relaxation times $T_1$ measured via exponential decay.
    \textbf{d}, Hahn-echo coherence times $T_{2}^{\mathrm{echo}}$ characterizing dephasing under echo refocusing.}
    \label{qubit_parameters}
\end{figure*}

The frequency allocation of the 30 working qubits is primarily constrained by the presence of two-level-system (TLS) defects, which introduce coherence dips and unstable frequency regions.  
To ensure reliable operation, idle frequencies are chosen to avoid TLS hotspots and other unstable intervals identified from spectroscopy and time-domain measurements.  
In addition, each qubit must remain sufficiently detuned from its neighbors to prevent frequency collisions during flux tuning.  
Because the entangling operations require pairs of qubits to be tuned toward a common interaction frequency, qubits forming two-qubit gates are assigned idle frequencies that are reasonably close.  
Finally, qubits sharing a common readout resonator must be well separated in frequency at their idle points to suppress measurement-induced crosstalk during simultaneous mid-circuit readout.

The resulting frequency arrangement and coherence parameters are summarized in Fig.~\ref{qubit_parameters}.  
Panel~(a) shows the assigned idle qubit frequencies, typically around $4.2$–$4.4~\mathrm{GHz}$.  
Panel~(b) displays the corresponding readout-resonator frequencies near $6.2~\mathrm{GHz}$.  
Panels~(c) and~(d) present the measured energy-relaxation times $T_1$ and Hahn-echo coherence times $T_2^{\mathrm{echo}}$, which characterize the coherence performance of the selected qubits.  
These measurements define the stable frequency manifold used for simultaneous mid-circuit measurements and high-fidelity quantum-gate operations.

\subsection{Gate implementation and calibration}

Single-qubit gates are implemented using Gaussian-DRAG microwave pulses~\cite{Motzoi2009,Chow2010}, with a nominal duration of $50~\mathrm{ns}$ for a $\pi$ rotation and $25~\mathrm{ns}$ for a $\pi/2$ rotation obtained by halving the pulse area.  
Two-qubit entangling operations are realized using tunable-coupler-mediated iSWAP-type interactions~\cite{Sung2021}, which are dressed with local single-qubit rotations to form the effective unitary layers used in the experiment; in practice, discrete equatorial $\pi/2$ pulses are applied before and after each entangling gate to control residual phases and generate locally randomized unitary dynamics.
For each iSWAP-type entangling gate, the participating qubits are brought into resonance through the coupler, and a parametric flux pulse is applied to activate the exchange interaction for a calibrated duration of $40~\mathrm{ns}$, corresponding to one full excitation-swap cycle at the operating point.

\begin{figure*}[t]
    \centering
    \includegraphics[width=1\textwidth]{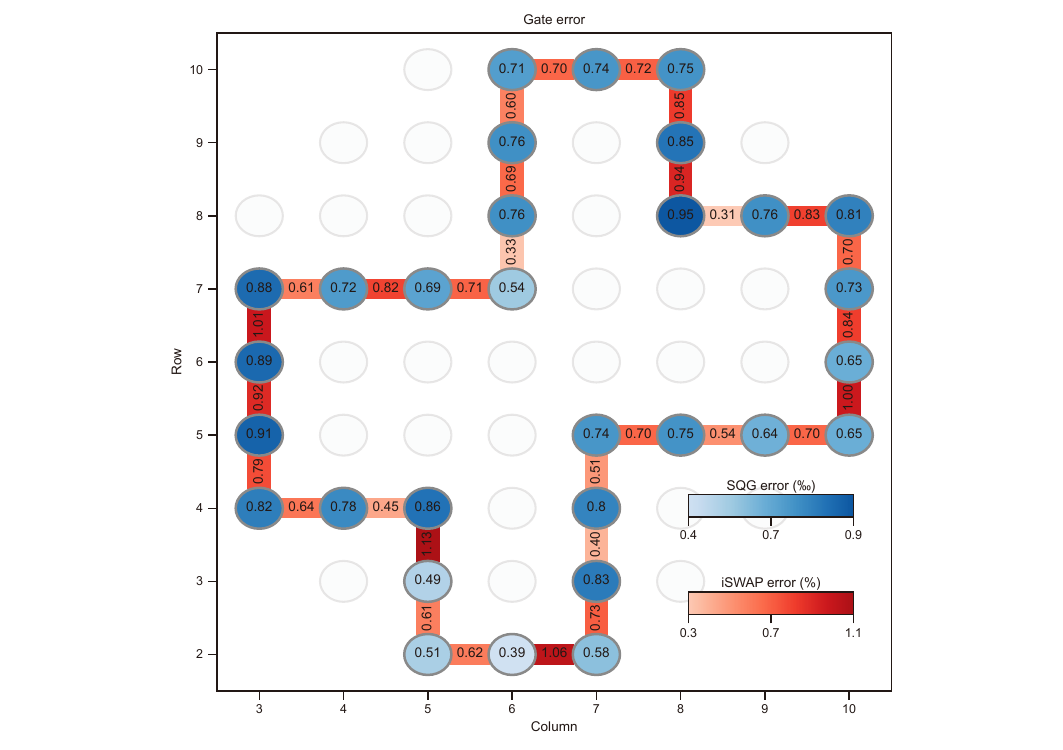}
    \caption{
    \textbf{Errors of single- and two-qubit gates.}
    Shown are the calibrated single-qubit gate errors and the corresponding two-qubit gate errors associated with the 30 qubits used in the experiment.
    }
    \label{Gate_error}
\end{figure*}

The gate fidelities are characterized using simultaneous benchmarking protocols.  
Single-qubit errors are extracted from simultaneous randomized benchmarking (RB)~\cite{Emerson2005} across all 30 qubits, yielding an average error of approximately $7\times 10^{-4}$.  
Two-qubit fidelities are obtained using simultaneous speckle purity benchmarking (SPB)~\cite{Arute2019}, in which odd and even couplers along the chain are benchmarked in alternating configurations to capture correlated errors.  
This procedure yields an average iSWAP-type gate error of about $7\times 10^{-3}$.  
The calibrated single- and two-qubit gate errors used in the experiment are summarized in Fig.~\ref{Gate_error}.

\clearpage
\bibliographystyle{naturemag}
\bibliography{supp}